\documentclass[aps,floatfix,prc,showpacs,showkeys,amsmath,amssymb,superscriptaddress]{revtex4-1}
\usepackage{graphicx,ctable}
\usepackage{amssymb}
\usepackage{hyperref}
\usepackage{float}
\usepackage{multirow}
\newcommand {\dNchdeta} {\ensuremath{\mathrm{d}N_\mathrm{ch}/\mathrm{d}\eta }}
\newcommand {\Tch} {\ensuremath{T_\mathrm{ch}}}
\begin{document}
\title{A Comparison of p-p, p-Pb, Pb-Pb Collisions in the Thermal Model:\\
Multiplicity Dependence of  Thermal Parameters.}
\author{Natasha Sharma}
\affiliation{Department of Physics, Panjab University, Chandigarh 160014, India}
\author{Jean Cleymans}
\affiliation{UCT-CERN Research Centre and Department of Physics, University of Cape Town, Rondebosch 7701, South Africa}
\author{Boris Hippolyte} 
\affiliation{Institut Pluridisciplinaire Hubert Curien and Universit\'e de Strasbourg Institute for Advanced Study, CNRS-IN2P3, Strasbourg, France}
\author{Masimba Paradza}
\affiliation{UCT-CERN Research Centre and Department of Physics, University of Cape Town, Rondebosch 7701, South Africa}
\date{30 October 2018}
\begin{abstract}
An analysis is made of the particle composition (hadrochemistry) of the final state in proton-proton (p-p), proton-lead (p-Pb)
and lead-lead (Pb-Pb) collisions
as a function of the charged particle multiplicity ($\dNchdeta$). The thermal model is used
to determine the chemical freeze-out temperature as well as the radius and strangeness saturation
factor $\gamma_s$. Three different ensembles are used in the analysis namely, the grand canonical ensemble,
the canonical ensemble with exact strangeness conservation and the canonical ensemble with exact
baryon number, strangeness and electric charge conservation. It is shown that for  high
multiplicities (at least 20 charged hadrons in the mid-rapidity interval considered) the three ensembles  lead to the same results. 
\end{abstract}
\pacs{12.40.Ee, 25.75.Dw}
\keywords{Thermal model, Strangeness, Particle production, Hadrochemistry}
\maketitle
%
\section{\label{secIntroduction}Introduction}
%
In high energy collisions applications of the thermal-statistical model in the form of the hadron resonance gas
model have been successful~(see e.g.~\cite{Andronic:2017pug,Becattini:2017pxe} for two recent
publications) in describing the composition of the final state e.g. the yields of pions,
kaons, protons and other hadrons. In these descriptions use is made of the grand canonical ensemble
and the canonical ensemble with exact strangeness conservation. In this paper we consider in addition
the use of the canonical ensemble with exact baryon, strangeness and charge conservation.  We also make
a systematic analysis of the dependence on the charged particle multiplicity \dNchdeta for the first time.

The identifying feature of the thermal model  is that all  hadronic resonances  listed in~\cite{Patrignani:2016xqp}  are
assumed to be in thermal and chemical equilibrium.
This  assumption drastically reduces the number of free parameters as this stage is determined by just a few
thermodynamic variables namely, the chemical freeze-out temperature $T_{ch}$, the various chemical potentials $\mu$ determined by
the conserved quantum numbers and by the volume $V$ of the system.
It has been shown that this description is also the correct
one~\cite{Cleymans:1997eq,Broniowski:2001we,Akkelin:2001wv} for a scaling expansion as first discussed by
Bjorken~\cite{Bjorken:1982qr}.  
After integration over $p_T$ these authors have shown that:
\begin{equation}
{dN_i/dy\over dN_j/dy} = {N_i^0\over N_j^0} 
\end{equation}
where $N^0_i$ ($N_j^0$)is the particle yield of hadron $i$ ($j$)
 as calculated in a fireball at rest, while $dN_i/dy$ is the yield of  hadron $i$  on the rapidity plateau.
Hence, in the Bjorken model with longitudinal scaling and radial expansion the effects of hydrodynamic flow cancel out in ratios.

The yields produced in heavy-ion collisions have been the subject of intense discussions over the past few years
and several proposals have been made in view of the fact that the number of pions is underestimated while the number of protons 
is overestimated. Several proposals to improve on this have been made recently:
\begin{itemize}
\item Incomplete hadron spectrum~\cite{Noronha-Hostler:2014aia},
\item chemical non-equilibrium at freeze-out~\cite{Petran:2013lja,Begun:2013nga,Begun:2014rsa},
\item modification of hadron abundances in the hadronic phase~\cite{Steinheimer:2012rd,Becattini:2012xb,Becattini:2016xct},
\item separate freeze-out for strange and non-strange hadrons~\cite{Chatterjee:2013yga,Bellwied:2013cta,Chatterjee:2016cog,Bellwied:2016kpj},
\item excluded volume interactions~\cite{ Alba:2016hwx},
\item  energy dependent Breit-Wigner widths~\cite{Vovchenko:2018fmh},
\item use the phase shift analysis to take into account repulsive and attractive interactions~\cite{Dash:2018mep,Andronic:2018qqt},
\item use the K-matrix formalism to take interactions into account~\cite{Dash:2018can}.
\end{itemize}
These proposals improve the agreement with the observed yields and furthermore, some  of them  change the chemical freeze-out
temperature, $T_{ch}$ in only a minimal way like those presented recently in~\cite{Vovchenko:2018fmh,Andronic:2018qqt}.  In the 
present analysis
we therefore kept to the basic structure of the thermal model 
with a single freeze-out temperature and focus on the resulting thermal parameters $T_{ch},\gamma_s$ and
the radius. 
All our  calculations 
were done using the latest version of THERMUS~\cite{Wheaton:2004qb}~\footnote{B. Hippolyte and Y. Schutz,
https://github.com/thermus-project/THERMUS}.

 Our results show some interesting  new features:
\begin{itemize}
\item the grand canonical ensemble, the ensemble with strict strangeness conservation and the one with strict
baryon number, strangeness and charge conservation agree very well for the particle composition in Pb-Pb collisions, 
they also agree well for p-Pb collisions
but  marked differences for p-p collisions are present. These differences  disappear as the multiplicity of charged particles
increases in the final state.  Thus, p-p collisions  with high multiplicities agree with what is
seen in large systems like p-Pb and  Pb-Pb collisions. Quantitatively this agreement starts when there are at least 
20 charged hadrons in the 
mid-rapidity interval being considered. 
It also throws doubt on the applicability of the thermal model as applied to p-p collisions with low multiplicity.
\item  The convergence of the results in the three ensembles lends support to the idea that  one reaches
a thermodynamic limit where the results are independent of the ensemble being used.
\end{itemize} 

\section{Ensembles considered in the thermal model}
We compare in great detail three different ensembles based on the thermal model.
\begin{itemize}
\item Grand canonical ensemble (GCE), the conservation of quantum numbers is implemented using chemical potentials. 
The quantum numbers are conserved on the average.
The partition function depends on thermodynamic quantities and the Hamiltonian describing the system of $N$ hadrons: 
\begin{equation}
Z_{GCE} = \textrm{Tr} \left[ e^{-(H-\mu N)/T}\right]
\end{equation}
which, in the framework of the thermal model considered here, leads to
\begin{equation}
\ln Z_{GCE}(T,\mu,V) = \sum_i g_i V \int\frac{d^3p}{(2\pi)^3}\exp\left( -\frac{E_i-\mu_i}{T} \right)\\
\end{equation}
in the Boltzmann approximation, $g_i$ is the degeneracy factor of hadron $i$, $V$ is the volume of the system, $\mu_i$
is the chemical potential associated with the hadron.
The yield is given by:
\begin{equation}
N_i^{GCE} = V \int \frac{d^3p}{(2\pi)^3} \exp \left( -\frac{E_i}{T}\right)  ,
\end{equation}
where we have put the chemical potentials equal to zero, as relevant for the beam energies at the Large Hadron
Collider considered here.
The decays of resonances have to be added to the final yield
\begin{equation}
N_i^{GCE}(\mathrm{total}) = N_i^{GCE} + \sum_j Br(j\rightarrow i) N_i^{GCE}  .
\end{equation}

\item Canonical ensemble with exact implementation of strangeness conservation, we will 
refer to this as the strangeness canonical ensemble (SCE). 
There are chemical potentials for baryon number $B$ and charge $Q$ but not for strangeness:
\begin{equation}
Z_{SCE} = \textrm{Tr}\left[ e^{-(H-\mu N)/T}\delta_{(S,\sum_iS_i)}\right]
\end{equation}
The delta function imposes exact strangeness conservation, requiring overall strangeness to be fixed to the value $S$, in this paper
we will only consider the case where overall strangeness is zero, $S=0$.
This change leads to~\cite{BraunMunzinger:2001as}:
\begin{equation}
Z_{SCE} = \frac{1}{(2\pi)}
\int_0^{2\pi} d\phi e^{-iS\phi}
Z_{GCE}(T,\mu_B,\lambda_S)
\end{equation}
where the fugacity factor is replaced by
\begin{equation}
\lambda_S = e^{i\phi}
\end{equation}
\begin{eqnarray}
N_{i}^{SCE}&=&V{{Z^1_{i}}\over {Z_{S=0}^C}} \sum_{k,p=-\infty}^{\infty} a_{3}^{p} a_{2}^{k}
a_{1}^{{-2k-3p- s}} I_k(x_2) I_p(x_3) I_{-2k-3p- s}(x_1),   \label{equ6}
\end{eqnarray}
where $Z^C_{S=0}$ is the canonical partition function
\begin{eqnarray}
Z^C_{S=0}&=&e^{S_0} \sum_{k,p=-\infty}^{\infty} a_{3}^{p}
a_{2}^{k} a_{1}^{{-2k-3p}}\nonumber 
 I_k(x_2) I_p(x_3) I_{-2k-3p}(x_1),
\label{eq7}
\end{eqnarray}
where $Z^1_i$ is the one-particle partition function calculated for $\mu_S=0$ in the Boltzmann
approximation. The arguments of the Bessel functions $I_s(x)$ and the parameters $a_i$ are introduced as,
\begin{eqnarray} a_s= \sqrt{{S_s}/{S_{\mathrm{-s}}}}~~,~~ x_s = 2V\sqrt{S_sS_{\mathrm{-s}}} \label{eq8a}, \end{eqnarray}
where $S_s$ is the sum  of all $Z^1_k(\mu_S=0)$  for particle species $k$ carrying strangeness
$s$. 
As previously, the decays of resonances have to be added to the final yield
\begin{equation}
N_i^{SCE}(\mathrm{total}) = N_i^{SCE} + \sum_j Br(j\rightarrow i) N_i^{SCE}  .
\end{equation}

\item Canonical ensemble with exact implementation of $B$, $S$ and $Q$ conservation, we will refer to this as the 
full canonical ensemble (FCE). 
In this ensemble 
there are no chemical potentials. The partition function is given by:
\begin{equation}
Z_{FCE} = \textrm{Tr}\left[ e^{-(H-\mu N)/T}\delta_{(B,\sum_iB_i)}\delta_{(Q,\sum_iQ_i)}\delta_{(S,\sum_iS_i)}\right]
\end{equation}
\begin{equation}
Z_{FCE} = \frac{1}{(2\pi)^3}
\int_0^{2\pi} d\alpha e^{-iB\alpha}
\int_0^{2\pi} d\psi e^{-iQ\psi}
\int_0^{2\pi} d\phi e^{-iS\phi}
Z_{GCE}(T,\lambda_B,\lambda_Q,\lambda_S)
\end{equation}
where the fugacity factors have been replaced by
\begin{equation}
\lambda_B = e^{i\alpha},\quad \lambda_Q = e^{i\psi}, \quad \lambda_S = e^{i\phi}   .
\end{equation}
As before, the decays of resonances have to be added to the final yield
\begin{equation}
N_i^{FCE}(\mathrm{total}) = N_i^{FCE} + \sum_j Br(j\rightarrow i) N_i^{FCE}  .
\end{equation}

\end{itemize}
In this case the analytic expression becomes very lengthy and we refrain from writing it down here, it is implemented in the 
THERMUS program~\cite{Wheaton:2004qb}.

In all three case we have also taken into account the strangeness saturation factor $\gamma_s$~\cite{Letessier:1993hi}
which enters as a multiplicative factor, raised to the power of the strangeness content,  in the particle yields.
Keeping this factor fixed at one does not change the fixed message, only the resulting value of $\chi^2$ is increased
indicating a worsening of the fits.

These three ensembles are applied to p-p collisions at 7 TeV in the central region of rapidity~\cite{ALICE:2017jyt},
to p-Pb collisions at 5.02 TeV~\cite{Abelev:2013haa,Adam:2015vsf} and to Pb-Pb collisions 
at 2.76 TeV~\cite{Abelev:2013vea,Abelev:2013xaa,Abelev:2013zaa}  with particular focus on the dependence on 
the charged particle multiplicity.
It is well known that in this kinematic region, 
one has particle - antiparticle symmetry and therefore there is no net baryon density and also no net strangeness. 
The different ensembles nevertheless
give different results because of the way they are implemented. A clear size dependence is present in the results of the ensembles.
In the thermodynamic limit they should become equivalent. 
Clearly there are other ensembles that could be investigated and also other sources of finite volume corrections. 
We hope to address these
in a future publication. 

A similar analysis was done in~\cite{Abelev:2006cs,Becattini:2009ee,Becattini:2010sk} for p-p collisions at 200 GeV but 
without the  dependence  on charged multiplicity.

For p-p collisions we have taken  the
five particle species listed in table~\ref{tab:cent-event} where we also compare the measured values with the model calculations.
For  p-Pb and Pb-Pb collisions we included the $\Omega$ measurements in our
analysis, so that   six particle species were considered  for p-Pb and Pb-Pb. 
We have checked explicitly that for the five bins in p-p collisions where the $\Omega$ has also been
measured, there is no difference in the outcome for the values of $T_{ch}, \gamma_s$ and the radius.

As shown in~\cite{Vislavicius:2016rwi,Sharma:2018owb} the $\phi$ meson 
is not described very well and has not been included.


\section{Comparison of different ensembles.}
\begin{figure}
\begin{center}
\includegraphics[width=0.7\textwidth,height=16cm]{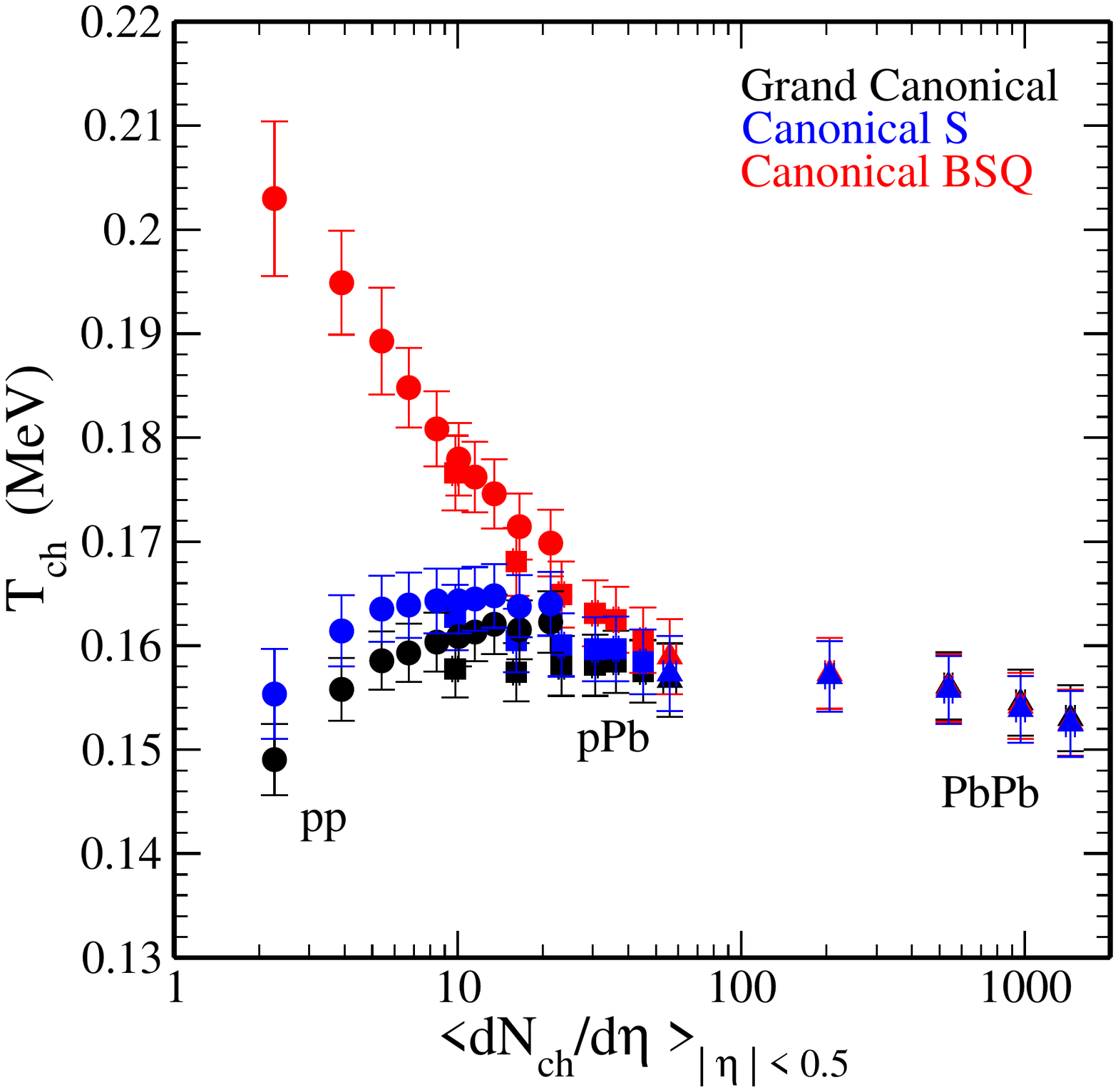}
\caption{\label{Tch} The chemical freeze-out temperature, $\Tch$, obtained for three different ensembles.
The black points are obtained using the grand canonical ensemble, the blue points use exact strangeness conservation
while the red points have built-in exact baryon number, strangeness and charge conservation.
Circles are for p-p collisions at 7 TeV, 
squares are for p-Pb collisions at 5.02 TeV while triangles are for Pb-Pb 
collisions at 2.76 TeV.
}
\end{center}
\end{figure}

%
%
In Fig.~\ref{Tch} we show the chemical freeze-out temperature as a function of the multiplicity of hadrons in the final state~\cite{ALICE:2017jyt}.
As explained in the previous section the freeze-out temperature has been calculated using three different ensembles. 
The highest values are obtained using the canonical ensemble
with exact conservation of three quantum numbers, baryon number $B$, strangeness $S$ and charge $Q$, all of them
being set to zero as is appropriate for the central rapidity region in p-p collisions at 7~TeV. 
In this ensemble the temperature drops 
strongly from the lowest to the highest multiplicity. 

\begin{table}[H]
  \centering
	\caption{\label{tab:cent-event}Comparison between measured and fitted values for p-p collisions at 7 TeV for V0M multiplicity class II.} 
	\vspace{0.1in}
	\begin{tabular}{|c|c|c|c|c|}
	\hline	
	\multirow{2}{*}{Particle Species} & \multirow{2}{*}{$dN/dy$ (data)} & \multicolumn{3}{c|}{$dN/dy$ (model)} \\
	\cline{3-5}
	&       &  Canonical S  &  Canonical B, S, Q & Grand Canonical  \\
		\hline
	$\pi^{\pm}$ & 7.88   $\pm$ 0.38   &  6.78 & 6.76 &   6.96 \\	
	\hline
	$K^0_S$ &  1.04    $\pm$ 0.05 & 1.16 & 1.16 & 1.15  \\
	\hline
	 $p, (\bar{p})$ &  0.44     $\pm$ 0.03  & 0.50 & 0.50 & 0.50\\
	\hline
	$\Lambda$ & 0.302     $\pm$  0.020 & 0.259 & 0.262 &  0.246 \\
	\hline
	$\Xi^{-}(\bar{\Xi}^+$) & 0.0358    $\pm$  0.0023  & 0.035 & 0.035  &  0.036 \\
	\hline
	\end{tabular}
\end{table}

The lowest values for $\Tch$ are obtained when using the grand canonical ensemble, in this case the conserved quantum numbers are 
again zero.
The results are clearly different from those obtained in the previous ensemble, especially in the low multiplicity intervals. They gradually approach
each other and they become equal at the highest multiplicities.

For comparison with the previous two cases we also calculated $\Tch$ using the canonical ensemble with only
strangeness $S$ being exactly conserved using the method presented in~\cite{BraunMunzinger:2001as}. In this case the results are 
 close to those 
obtained in the grand canonical ensemble, with the values of $\Tch$ always slightly higher than in the grand canonical ensemble. Again for the 
highest multiplicity interval the results become equivalent.
As can be seen in  Fig.~\ref{Tch}, even though all the ensembles produce different results, for high multiplicities the results 
converge to a common value close to 160 MeV.

\begin{figure}
\begin{center}
\includegraphics[width=0.7\textwidth,height=16cm]{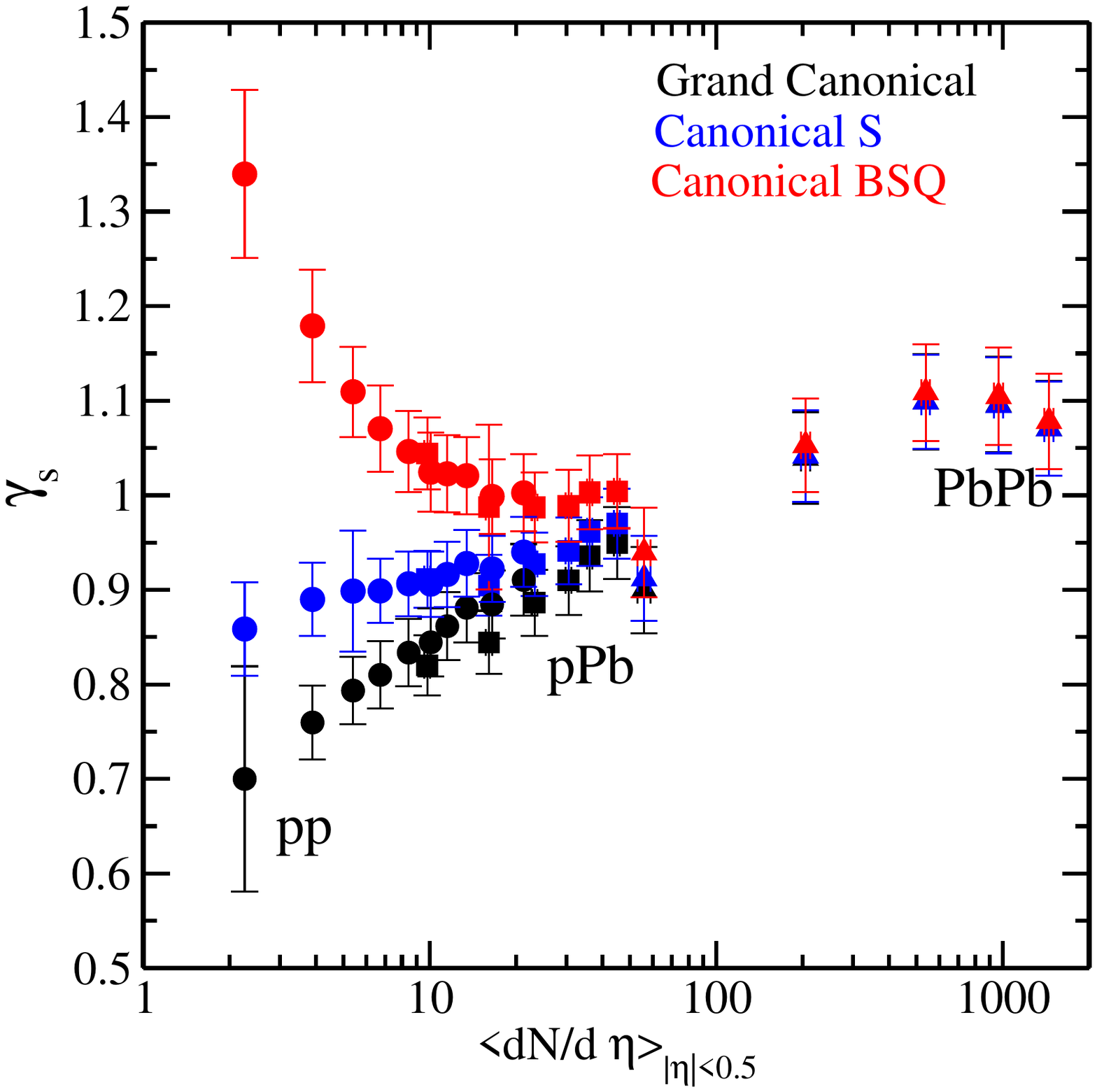}
\caption{\label{gammas} The strangeness saturation factor $\gamma_s$ obtained for three different ensembles.  
The black points were obtained using the grand canonical ensemble, the blue points uses exact strangeness conservation
while the red points have built-in exact baryon number, strangeness and charge conservation.
Circles are for p-p collisions at 7 TeV, 
squares are for p-Pb collisions at 5.02 TeV while triangles are for Pb-Pb 
collisions at 2.76 TeV.
}
\end{center}
\end{figure}
In Fig.~\ref{gammas}  we show results for the strangeness saturation factor $\gamma_s$~\cite{Letessier:1993hi}. 
In this case we obtain again quite substantial differences in each
one of the three ensembles considered. The highest values being found in the canonical ensemble with exact strangeness conservation.
Note that the values of $\gamma_s$ become compatible with unity, i.e.  with chemical equilibrium for all light flavors.

\begin{figure}
\begin{center}
\includegraphics[width=0.7\textwidth,height=16cm]{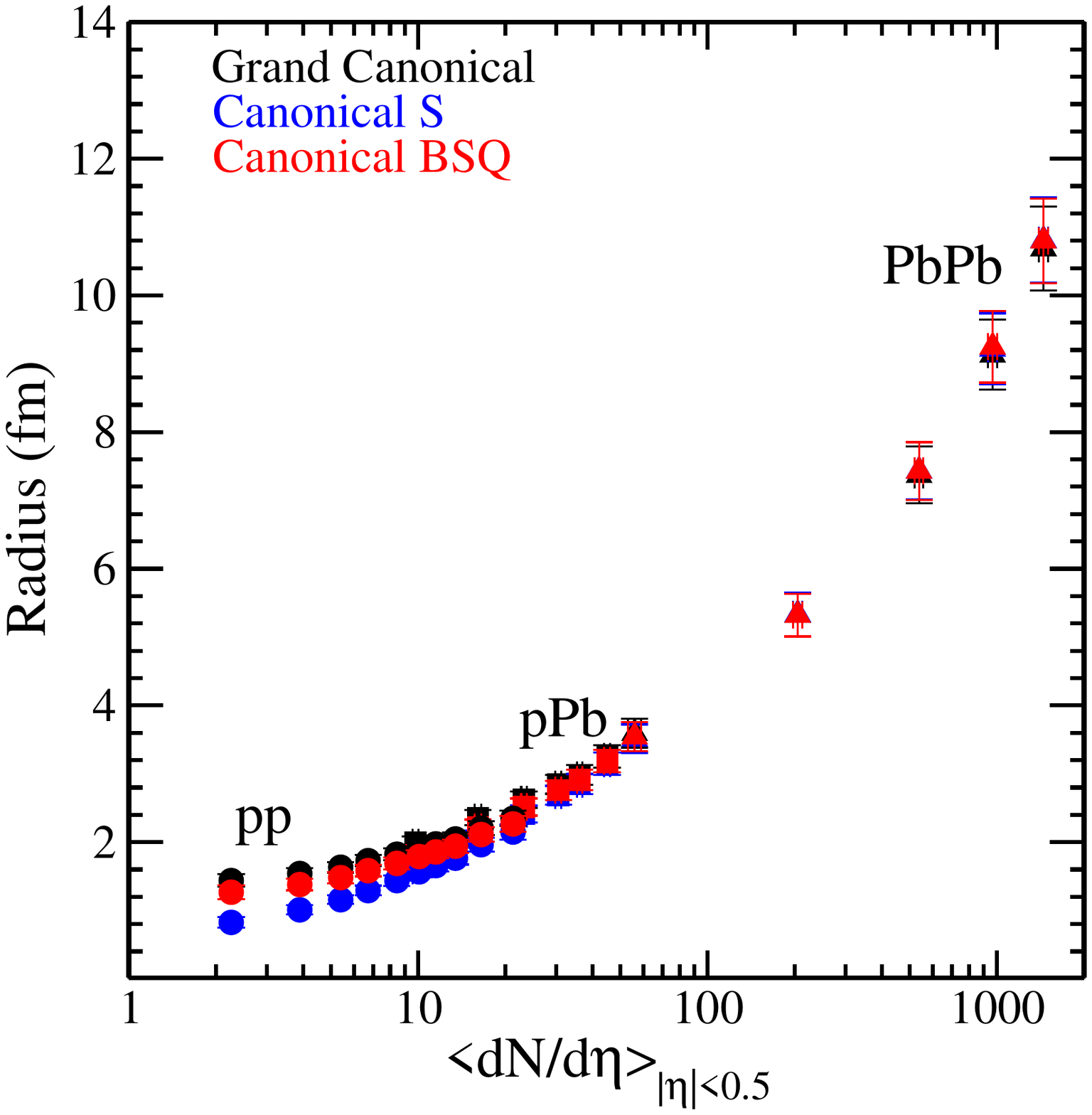}
\caption{\label{radius}The chemical freeze-out radius obtained for three different ensembles.  
The black points were obtained using the grand canonical ensemble, the blue points uses exact strangeness conservation
while the red points have built-in exact baryon number, strangeness and charge conservation.
Circles are for p-p collisions at 7 TeV, 
squares are for p-Pb collisions at 5.02 TeV while triangles are for Pb-Pb 
collisions at 2.76 TeV.
}
\end{center}
\end{figure}
In Fig.~\ref{radius} the radius at chemical  freeze-out obtained in the three ensembles is presented. As in the previous figures, the results become 
independent of the ensemble chosen for the highest multiplicities while showing clear differences for low multiplicities.

Our results  show that there is a strong correlation between some of the parameters. 
The very high temperature obtained in the canonical $BSQ$
ensemble (FCE) correlates with the small radius in the same ensemble. Particle yields increase with temperature but a small volume
decreases them, hence the correlation between the parameters.

Table 2 shows the $\chi^2$ values obtained for the three ensembles considered in this paper.
The values confirm the earlier statement that these values  throw doubt on the applicability of the thermal model in
p-p collisions.
\begin{table}[ht]
\begin{center}
\begin{tabular}{|c|c|c|c|}
\hline
$\langle dN_{ch}/d\eta \rangle|_{|\eta|<0.5}$& Canonical S  & Canonical B, S, Q & Grand Canonical      \\
\hline 
\multicolumn{4}{|c|}{p-p collisions}\\
\hline
2.26  &   12.79 / 2 &  3.85  / 2  & 6.45 / 2 \\
3.9   &   20.16 / 2 &  9.15  / 2 & 14.47 / 2 \\
5.4   &  25.46  / 2 & 14.94 / 2 & 20.27 / 2 \\
6.72  &  24.61  / 2 & 16.58 / 2 & 20.09 / 2 \\
8.45  &  24.65  / 2 & 18.71 / 2 & 20.83 / 2 \\
10.08 &  24.45  / 2 & 20.03 / 2 & 21.61 / 2 \\
11.51 &  24.42  / 2 & 20.91 / 2 & 21.80 / 2 \\
13.46 &  24.84  / 2 & 22.25 / 2 & 22.46 / 2 \\
16.51 &  23.52  / 2 & 22.19 / 2 & 22.41 / 2 \\
21.29 &  22.20  / 2 & 21.83 / 2 & 21.55 / 2 \\
\hline
\multicolumn{4}{|c|}{p-Pb collisions}\\
\hline
9.8     &      22.37 / 3&    17.86 /  3    &  18.72 / 3	     \\
16.1 	&   20.07 / 3	&    19.36 /  3    &  21.10 / 3	     \\
23.2 	&   19.23 / 3	&    19.71 /  3    &  20.60 / 3	     \\
30.5 	&    18.23 / 3	&    19.23 /  3    &  19,76 / 3	     \\
36.2   	&     18.17 / 3	&    19.39 /  3    &  19.42 / 3	     \\
45     	&     18.43 / 3	&    19.81 /  3    &  20.26 / 3	     \\
\hline
\multicolumn{4}{|c|}{Pb-Pb collisions}\\
\hline
56    &	 12.05 / 23    &   13.32 / 3    & 12.24 /  3          \\
205   &	 12.89 / 23    &   14.72 / 3    & 13.18 /  3          \\
538   &  15.54 / 23    &   17.94 / 3    & 16.14 /  3          \\
966   &	 13.48 / 23    &   16.07 / 3    &      14.26 /  3      \\
1448  &  11.08 / 23    &   13.07 / 3    &      11.45  / 3	  \\
\hline                                           
\end{tabular}
\caption{Values of $\chi^2$/ndf for various fits. 
}
\end{center}
\end{table}
%
%
\begin{figure}
\begin{center}
\includegraphics[width=\textwidth,height=16cm]{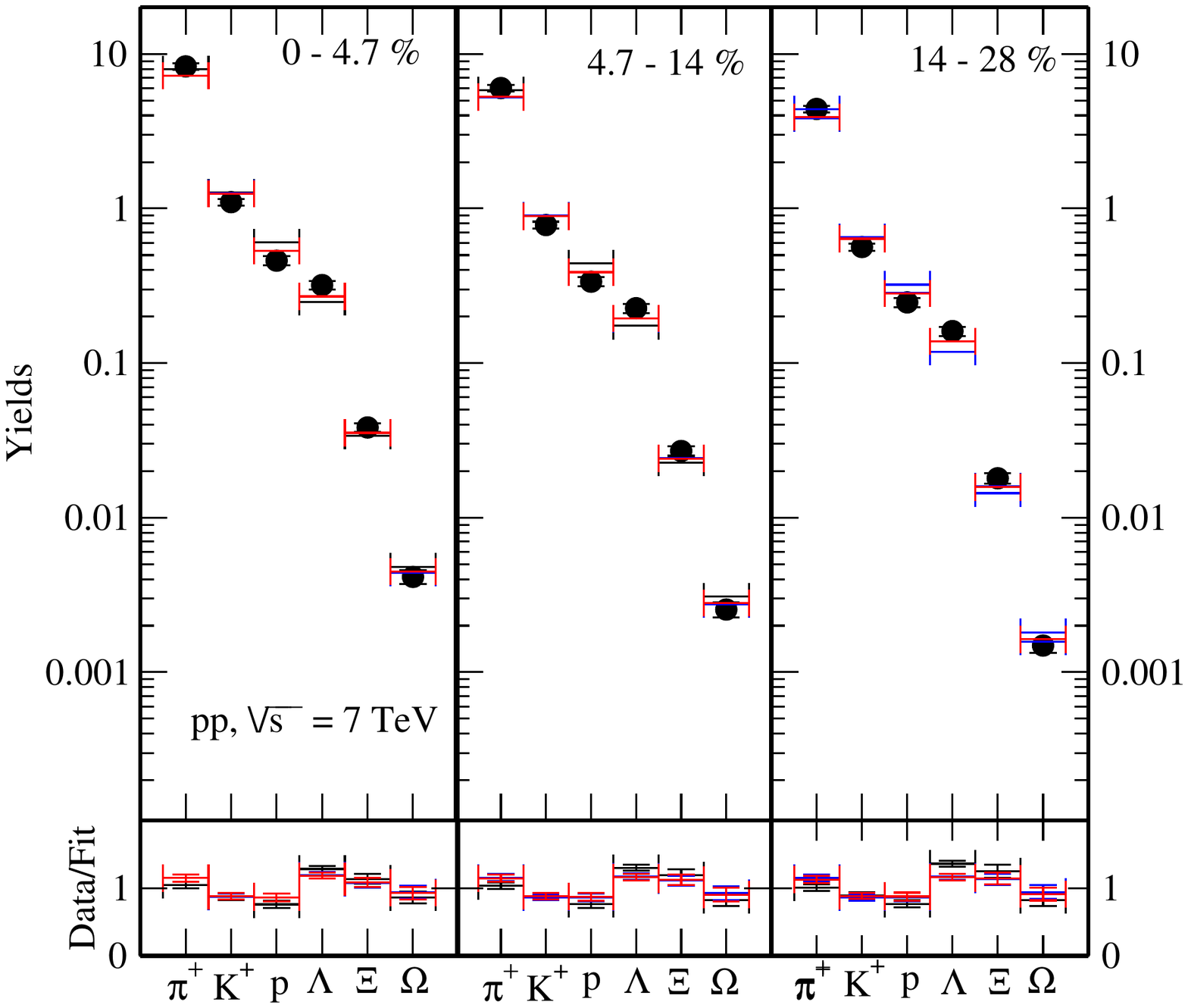}
\caption{Hadronic yields in p-p collisions at 7 Tev in centrality bins  1, 2 and 3 corresponding to 0 - 4.7 \% 
(left panel) and 4.7 - 14 \%  (middle panel)  and  14 - 28 \%(right panel)
respectively. The lower panel shows the ratio of experimental data divided by the fit results.
The black points were obtained using the grand canonical ensemble, the blue points uses exact strangeness conservation
while the red ones have built-in exact baryon number, strangeness and charge conservation. In some cases the bars overlap.
The data are taken  from Ref.~\cite{ALICE:2017jyt}.}
\end{center}
\label{cent123}
\end{figure}
\begin{figure}
\begin{center}
\includegraphics[width=\textwidth,height=16cm]{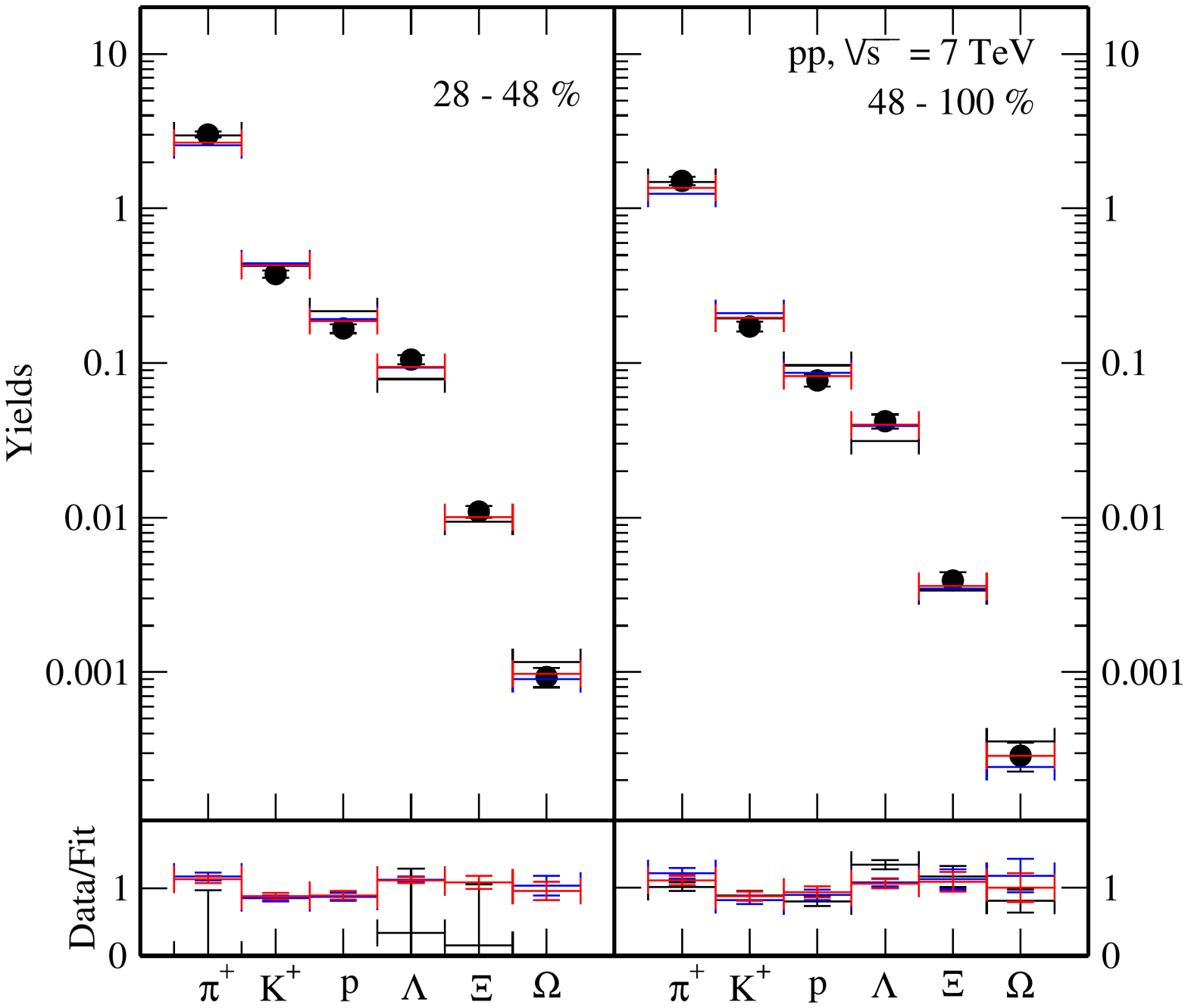}
\caption{Hadronic yields in p-p collisions at 7 Tev in centrality bins  4 and 5 corresponding to 28 - 48 \% 
(left panel) and 48 - 100 \%  (right panel) 
respectively. The lower panel shows the ratio of experimental data divided by the fit results.
The black bars were obtained using the grand canonical ensemble, the blue ones uses exact strangeness conservation
while the red ones have been obtained using exact baryon number, strangeness and charge conservation.
The data are taken  from Ref.~\cite{ALICE:2017jyt}.
}
\end{center}
\label{cent45}
\end{figure}
Fixing $\gamma_s$ = 1 does not change the physics message but considerably worsens the resulting $\chi^2$ values.

The fits to the hadronic yields obtained in p-p collisions at 7 TeV in five different
centrality bins  are shown in Figs~4 and 5.  The upper panels show the yields while the lower panels
show the ratios of the measured data divided by the fit values for the three different ensembles considered here.
The three lines corresponding to the fits are often very close to each other and overlap, hence they are not always 
visible on the figures.

\section{Discussion and Conclusions}
%
%
In this paper we have investigated three different ensembles to analyze 
the variation of particle yields with the 
multiplicity of charged particles produced in proton-proton collisions at the center-of-mass energy
of $\sqrt{s}$~=~7~TeV~\cite{ALICE:2017jyt}, p-Pb collisions at 5.02 TeV~\cite{Abelev:2013haa,Adam:2015vsf} and Pb-Pb collisions 
at 2.76 TeV~\cite{Abelev:2013vea,Abelev:2013xaa,Abelev:2013zaa}. 

We have kept the basic structure of the thermal model as presented in~\cite{Wheaton:2004qb} and focused on 
the resulting thermal parameters $T_{ch},\gamma_s$ and
the radius and their dependence on the final state multiplicity. We note in this regards that recent improvements on the treatment of 
the particle yields do not lead to substantial changes
of the chemical freeze-out temperature, $T_{ch}$~\cite{Vovchenko:2018fmh,Andronic:2018qqt}.  
Our results show two new interesting features:
\begin{itemize}
\item a comparison of the grand canonical ensemble, the ensemble with strict strangeness conservation and the one with strict
baryon number, strangeness and charge conservation agree very well for  large systems like p-Pb and Pb-Pb, 
but show marked differences for p-p collisions. These differences tend to disappear as the multiplicity of charged particles
increases in the final state of p-p collisions.  This supports the fact that   p-p collisions  with high multiplicities agree 
with what is
seen in large systems like Pb-Pb. Quantitatively this starts happening when there are more than 20 charged hadrons in the 
mid-rapidity interval being considered. 
It also throws doubt on the applicability of the thermal model in low multiplicity
p-p collisions.
\item  The convergence of the results in the three ensembles lends support to the notion a thermodynamic limit 
is reached where  results are independent of the ensemble being used.
\end{itemize}

We believe that it is of  interest to note that all three ensembles lead to the same results when the 
multiplicity of charged particles $\dNchdeta$ exceeds  20 at mid-rapidity. 
This could be interpreted as reaching the thermodynamic limit
since the three  ensembles lead to the same results.  
It would be of interest to extend this analysis to higher beam energies and higher multiplicity intervals.

\begin{acknowledgments}
\normalsize
One of us (J.C.) gratefully thanks  the National Research Foundation of South Africa for financial support.
N.S. acknowledges the support of SERB Ramanujan Fellowship
(D.O. No. SB/S2/RJN- 084/2015) of the Department of Science and Technology of India.
B.H. acknowledges the support of the Universit\'e de Strasbourg Institute for Advanced Study. 
\end{acknowledgments}
\newpage
\bibliographystyle{apsrev4-1}
\bibliography{sch_pp}

\begin{thebibliography}{38}%
\makeatletter
\providecommand \@ifxundefined [1]{%
 \@ifx{#1\undefined}
}%
\providecommand \@ifnum [1]{%
 \ifnum #1\expandafter \@firstoftwo
 \else \expandafter \@secondoftwo
 \fi
}%
\providecommand \@ifx [1]{%
 \ifx #1\expandafter \@firstoftwo
 \else \expandafter \@secondoftwo
 \fi
}%
\providecommand \natexlab [1]{#1}%
\providecommand \enquote  [1]{``#1''}%
\providecommand \bibnamefont  [1]{#1}%
\providecommand \bibfnamefont [1]{#1}%
\providecommand \citenamefont [1]{#1}%
\providecommand \href@noop [0]{\@secondoftwo}%
\providecommand \href [0]{\begingroup \@sanitize@url \@href}%
\providecommand \@href[1]{\@@startlink{#1}\@@href}%
\providecommand \@@href[1]{\endgroup#1\@@endlink}%
\providecommand \@sanitize@url [0]{\catcode `\\12\catcode `\$12\catcode
  `\&12\catcode `\#12\catcode `\^12\catcode `\_12\catcode `\%12\relax}%
\providecommand \@@startlink[1]{}%
\providecommand \@@endlink[0]{}%
\providecommand \url  [0]{\begingroup\@sanitize@url \@url }%
\providecommand \@url [1]{\endgroup\@href {#1}{\urlprefix }}%
\providecommand \urlprefix  [0]{URL }%
\providecommand \Eprint [0]{\href }%
\providecommand \doibase [0]{http://dx.doi.org/}%
\providecommand \selectlanguage [0]{\@gobble}%
\providecommand \bibinfo  [0]{\@secondoftwo}%
\providecommand \bibfield  [0]{\@secondoftwo}%
\providecommand \translation [1]{[#1]}%
\providecommand \BibitemOpen [0]{}%
\providecommand \bibitemStop [0]{}%
\providecommand \bibitemNoStop [0]{.\EOS\space}%
\providecommand \EOS [0]{\spacefactor3000\relax}%
\providecommand \BibitemShut  [1]{\csname bibitem#1\endcsname}%
\let\auto@bib@innerbib\@empty
\bibitem [{\citenamefont {Andronic}\ \emph {et~al.}(2017)\citenamefont
  {Andronic}, \citenamefont {Braun-Munzinger}, \citenamefont {Redlich},\ and\
  \citenamefont {Stachel}}]{Andronic:2017pug}%
  \BibitemOpen
  \bibfield  {author} {\bibinfo {author} {\bibfnamefont {A.}~\bibnamefont
  {Andronic}}, \bibinfo {author} {\bibfnamefont {P.}~\bibnamefont
  {Braun-Munzinger}}, \bibinfo {author} {\bibfnamefont {K.}~\bibnamefont
  {Redlich}}, \ and\ \bibinfo {author} {\bibfnamefont {J.}~\bibnamefont
  {Stachel}},\ }\href@noop {} {\  (\bibinfo {year} {2017})},\ \Eprint
  {http://arxiv.org/abs/nucl-th/1710.09425} {arXiv:nucl-th/1710.09425
  [nucl-th]} \BibitemShut {NoStop}%
\bibitem [{\citenamefont {Becattini}\ \emph
  {et~al.}(2017{\natexlab{a}})\citenamefont {Becattini}, \citenamefont
  {Bleicher}, \citenamefont {Steinheimer},\ and\ \citenamefont
  {Stock}}]{Becattini:2017pxe}%
  \BibitemOpen
  \bibfield  {author} {\bibinfo {author} {\bibfnamefont {F.}~\bibnamefont
  {Becattini}}, \bibinfo {author} {\bibfnamefont {M.}~\bibnamefont {Bleicher}},
  \bibinfo {author} {\bibfnamefont {J.}~\bibnamefont {Steinheimer}}, \ and\
  \bibinfo {author} {\bibfnamefont {R.}~\bibnamefont {Stock}},\ }\href@noop {}
  {\  (\bibinfo {year} {2017}{\natexlab{a}})},\ \Eprint
  {http://arxiv.org/abs/hep-ph/1712.03748} {arXiv:hep-ph/1712.03748 [hep-ph]}
  \BibitemShut {NoStop}%
\bibitem [{\citenamefont {Patrignani}\ \emph {et~al.}(2016)\citenamefont
  {Patrignani} \emph {et~al.}}]{Patrignani:2016xqp}%
  \BibitemOpen
  \bibfield  {author} {\bibinfo {author} {\bibfnamefont {C.}~\bibnamefont
  {Patrignani}} \emph {et~al.} (\bibinfo {collaboration} {Particle Data
  Group}),\ }\href {\doibase 10.1088/1674-1137/40/10/100001} {\bibfield
  {journal} {\bibinfo  {journal} {Chin. Phys.}\ }\textbf {\bibinfo {volume}
  {C40}},\ \bibinfo {pages} {100001} (\bibinfo {year} {2016})}\BibitemShut
  {NoStop}%
\bibitem [{\citenamefont {Cleymans}(1998)}]{Cleymans:1997eq}%
  \BibitemOpen
  \bibfield  {author} {\bibinfo {author} {\bibfnamefont {J.}~\bibnamefont
  {Cleymans}},\ }in\ \href@noop {} {\emph {\bibinfo {booktitle} {{Proceedings,
  3rd International Conference on Physics and astrophysics of quark-gluon
  plasma (ICPA-QGP '97)}}}}\ (\bibinfo {year} {1998})\ pp.\ \bibinfo {pages}
  {55--64},\ \Eprint {http://arxiv.org/abs/nucl-th/9704046}
  {arXiv:nucl-th/9704046 [nucl-th]} \BibitemShut {NoStop}%
\bibitem [{\citenamefont {Broniowski}\ and\ \citenamefont
  {Florkowski}(2001)}]{Broniowski:2001we}%
  \BibitemOpen
  \bibfield  {author} {\bibinfo {author} {\bibfnamefont {W.}~\bibnamefont
  {Broniowski}}\ and\ \bibinfo {author} {\bibfnamefont {W.}~\bibnamefont
  {Florkowski}},\ }\href {\doibase 10.1103/PhysRevLett.87.272302} {\bibfield
  {journal} {\bibinfo  {journal} {Phys. Rev. Lett.}\ }\textbf {\bibinfo
  {volume} {87}},\ \bibinfo {pages} {272302} (\bibinfo {year} {2001})},\
  \Eprint {http://arxiv.org/abs/nucl-th/0106050} {arXiv:nucl-th/0106050
  [nucl-th]} \BibitemShut {NoStop}%
\bibitem [{\citenamefont {Akkelin}\ \emph {et~al.}(2002)\citenamefont
  {Akkelin}, \citenamefont {Braun-Munzinger},\ and\ \citenamefont
  {Sinyukov}}]{Akkelin:2001wv}%
  \BibitemOpen
  \bibfield  {author} {\bibinfo {author} {\bibfnamefont {S.~V.}\ \bibnamefont
  {Akkelin}}, \bibinfo {author} {\bibfnamefont {P.}~\bibnamefont
  {Braun-Munzinger}}, \ and\ \bibinfo {author} {\bibfnamefont {{\relax
  Yu}.~M.}\ \bibnamefont {Sinyukov}},\ }\href {\doibase
  10.1016/S0375-9474(02)01165-X} {\bibfield  {journal} {\bibinfo  {journal}
  {Nucl. Phys.}\ }\textbf {\bibinfo {volume} {A710}},\ \bibinfo {pages} {439}
  (\bibinfo {year} {2002})},\ \Eprint {http://arxiv.org/abs/nucl-th/0111050}
  {arXiv:nucl-th/0111050 [nucl-th]} \BibitemShut {NoStop}%
\bibitem [{\citenamefont {Bjorken}(1983)}]{Bjorken:1982qr}%
  \BibitemOpen
  \bibfield  {author} {\bibinfo {author} {\bibfnamefont {J.~D.}\ \bibnamefont
  {Bjorken}},\ }\href {\doibase 10.1103/PhysRevD.27.140} {\bibfield  {journal}
  {\bibinfo  {journal} {Phys. Rev.}\ }\textbf {\bibinfo {volume} {D27}},\
  \bibinfo {pages} {140} (\bibinfo {year} {1983})}\BibitemShut {NoStop}%
\bibitem [{\citenamefont {Noronha-Hostler}\ and\ \citenamefont
  {Greiner}(2014)}]{Noronha-Hostler:2014aia}%
  \BibitemOpen
  \bibfield  {author} {\bibinfo {author} {\bibfnamefont {J.}~\bibnamefont
  {Noronha-Hostler}}\ and\ \bibinfo {author} {\bibfnamefont {C.}~\bibnamefont
  {Greiner}},\ }\bibfield  {booktitle} {\emph {\bibinfo {booktitle}
  {{Proceedings, 24th International Conference on Ultra-Relativistic
  Nucleus-Nucleus Collisions (Quark Matter 2014): Darmstadt, Germany, May
  19-24, 2014}}},\ }\href {\doibase 10.1016/j.nuclphysa.2014.08.101} {\bibfield
   {journal} {\bibinfo  {journal} {Nucl. Phys.}\ }\textbf {\bibinfo {volume}
  {A931}},\ \bibinfo {pages} {1108} (\bibinfo {year} {2014})},\ \Eprint
  {http://arxiv.org/abs/1408.0761} {arXiv:1408.0761 [nucl-th]} \BibitemShut
  {NoStop}%
\bibitem [{\citenamefont {Petr\'an}\ \emph {et~al.}(2013)\citenamefont
  {Petr\'an}, \citenamefont {Letessier}, \citenamefont {Petr\'a\v{c}ek},\ and\
  \citenamefont {Rafelski}}]{Petran:2013lja}%
  \BibitemOpen
  \bibfield  {author} {\bibinfo {author} {\bibfnamefont {M.}~\bibnamefont
  {Petr\'an}}, \bibinfo {author} {\bibfnamefont {J.}~\bibnamefont {Letessier}},
  \bibinfo {author} {\bibfnamefont {V.}~\bibnamefont {Petr\'a\v{c}ek}}, \ and\
  \bibinfo {author} {\bibfnamefont {J.}~\bibnamefont {Rafelski}},\ }\href
  {\doibase 10.1103/PhysRevC.88.034907} {\bibfield  {journal} {\bibinfo
  {journal} {Phys. Rev.}\ }\textbf {\bibinfo {volume} {C88}},\ \bibinfo {pages}
  {034907} (\bibinfo {year} {2013})},\ \Eprint {http://arxiv.org/abs/1303.2098}
  {arXiv:1303.2098 [hep-ph]} \BibitemShut {NoStop}%
\bibitem [{\citenamefont {Begun}\ \emph
  {et~al.}(2014{\natexlab{a}})\citenamefont {Begun}, \citenamefont
  {Florkowski},\ and\ \citenamefont {Rybczynski}}]{Begun:2013nga}%
  \BibitemOpen
  \bibfield  {author} {\bibinfo {author} {\bibfnamefont {V.}~\bibnamefont
  {Begun}}, \bibinfo {author} {\bibfnamefont {W.}~\bibnamefont {Florkowski}}, \
  and\ \bibinfo {author} {\bibfnamefont {M.}~\bibnamefont {Rybczynski}},\
  }\href {\doibase 10.1103/PhysRevC.90.014906} {\bibfield  {journal} {\bibinfo
  {journal} {Phys. Rev.}\ }\textbf {\bibinfo {volume} {C90}},\ \bibinfo {pages}
  {014906} (\bibinfo {year} {2014}{\natexlab{a}})},\ \Eprint
  {http://arxiv.org/abs/1312.1487} {arXiv:1312.1487 [nucl-th]} \BibitemShut
  {NoStop}%
\bibitem [{\citenamefont {Begun}\ \emph
  {et~al.}(2014{\natexlab{b}})\citenamefont {Begun}, \citenamefont
  {Florkowski},\ and\ \citenamefont {Rybczynski}}]{Begun:2014rsa}%
  \BibitemOpen
  \bibfield  {author} {\bibinfo {author} {\bibfnamefont {V.}~\bibnamefont
  {Begun}}, \bibinfo {author} {\bibfnamefont {W.}~\bibnamefont {Florkowski}}, \
  and\ \bibinfo {author} {\bibfnamefont {M.}~\bibnamefont {Rybczynski}},\
  }\href {\doibase 10.1103/PhysRevC.90.054912} {\bibfield  {journal} {\bibinfo
  {journal} {Phys. Rev.}\ }\textbf {\bibinfo {volume} {C90}},\ \bibinfo {pages}
  {054912} (\bibinfo {year} {2014}{\natexlab{b}})},\ \Eprint
  {http://arxiv.org/abs/1405.7252} {arXiv:1405.7252 [hep-ph]} \BibitemShut
  {NoStop}%
\bibitem [{\citenamefont {Steinheimer}\ \emph {et~al.}(2013)\citenamefont
  {Steinheimer}, \citenamefont {Aichelin},\ and\ \citenamefont
  {Bleicher}}]{Steinheimer:2012rd}%
  \BibitemOpen
  \bibfield  {author} {\bibinfo {author} {\bibfnamefont {J.}~\bibnamefont
  {Steinheimer}}, \bibinfo {author} {\bibfnamefont {J.}~\bibnamefont
  {Aichelin}}, \ and\ \bibinfo {author} {\bibfnamefont {M.}~\bibnamefont
  {Bleicher}},\ }\href {\doibase 10.1103/PhysRevLett.110.042501} {\bibfield
  {journal} {\bibinfo  {journal} {Phys. Rev. Lett.}\ }\textbf {\bibinfo
  {volume} {110}},\ \bibinfo {pages} {042501} (\bibinfo {year} {2013})},\
  \Eprint {http://arxiv.org/abs/1203.5302} {arXiv:1203.5302 [nucl-th]}
  \BibitemShut {NoStop}%
\bibitem [{\citenamefont {Becattini}\ \emph {et~al.}(2013)\citenamefont
  {Becattini}, \citenamefont {Bleicher}, \citenamefont {Kollegger},
  \citenamefont {Schuster}, \citenamefont {Steinheimer},\ and\ \citenamefont
  {Stock}}]{Becattini:2012xb}%
  \BibitemOpen
  \bibfield  {author} {\bibinfo {author} {\bibfnamefont {F.}~\bibnamefont
  {Becattini}}, \bibinfo {author} {\bibfnamefont {M.}~\bibnamefont {Bleicher}},
  \bibinfo {author} {\bibfnamefont {T.}~\bibnamefont {Kollegger}}, \bibinfo
  {author} {\bibfnamefont {T.}~\bibnamefont {Schuster}}, \bibinfo {author}
  {\bibfnamefont {J.}~\bibnamefont {Steinheimer}}, \ and\ \bibinfo {author}
  {\bibfnamefont {R.}~\bibnamefont {Stock}},\ }\href {\doibase
  10.1103/PhysRevLett.111.082302} {\bibfield  {journal} {\bibinfo  {journal}
  {Phys. Rev. Lett.}\ }\textbf {\bibinfo {volume} {111}},\ \bibinfo {pages}
  {082302} (\bibinfo {year} {2013})},\ \Eprint {http://arxiv.org/abs/1212.2431}
  {arXiv:1212.2431 [nucl-th]} \BibitemShut {NoStop}%
\bibitem [{\citenamefont {Becattini}\ \emph
  {et~al.}(2017{\natexlab{b}})\citenamefont {Becattini}, \citenamefont
  {Steinheimer}, \citenamefont {Stock},\ and\ \citenamefont
  {Bleicher}}]{Becattini:2016xct}%
  \BibitemOpen
  \bibfield  {author} {\bibinfo {author} {\bibfnamefont {F.}~\bibnamefont
  {Becattini}}, \bibinfo {author} {\bibfnamefont {J.}~\bibnamefont
  {Steinheimer}}, \bibinfo {author} {\bibfnamefont {R.}~\bibnamefont {Stock}},
  \ and\ \bibinfo {author} {\bibfnamefont {M.}~\bibnamefont {Bleicher}},\
  }\href {\doibase 10.1016/j.physletb.2016.11.033} {\bibfield  {journal}
  {\bibinfo  {journal} {Phys. Lett.}\ }\textbf {\bibinfo {volume} {B764}},\
  \bibinfo {pages} {241} (\bibinfo {year} {2017}{\natexlab{b}})},\ \Eprint
  {http://arxiv.org/abs/1605.09694} {arXiv:1605.09694 [nucl-th]} \BibitemShut
  {NoStop}%
\bibitem [{\citenamefont {Chatterjee}\ \emph {et~al.}(2013)\citenamefont
  {Chatterjee}, \citenamefont {Godbole},\ and\ \citenamefont
  {Gupta}}]{Chatterjee:2013yga}%
  \BibitemOpen
  \bibfield  {author} {\bibinfo {author} {\bibfnamefont {S.}~\bibnamefont
  {Chatterjee}}, \bibinfo {author} {\bibfnamefont {R.~M.}\ \bibnamefont
  {Godbole}}, \ and\ \bibinfo {author} {\bibfnamefont {S.}~\bibnamefont
  {Gupta}},\ }\href {\doibase 10.1016/j.physletb.2013.11.008} {\bibfield
  {journal} {\bibinfo  {journal} {Phys. Lett.}\ }\textbf {\bibinfo {volume}
  {B727}},\ \bibinfo {pages} {554} (\bibinfo {year} {2013})},\ \Eprint
  {http://arxiv.org/abs/1306.2006} {arXiv:1306.2006 [nucl-th]} \BibitemShut
  {NoStop}%
\bibitem [{\citenamefont {Bellwied}\ \emph {et~al.}(2013)\citenamefont
  {Bellwied}, \citenamefont {Borsanyi}, \citenamefont {Fodor}, \citenamefont
  {Katz},\ and\ \citenamefont {Ratti}}]{Bellwied:2013cta}%
  \BibitemOpen
  \bibfield  {author} {\bibinfo {author} {\bibfnamefont {R.}~\bibnamefont
  {Bellwied}}, \bibinfo {author} {\bibfnamefont {S.}~\bibnamefont {Borsanyi}},
  \bibinfo {author} {\bibfnamefont {Z.}~\bibnamefont {Fodor}}, \bibinfo
  {author} {\bibfnamefont {S.~D.}\ \bibnamefont {Katz}}, \ and\ \bibinfo
  {author} {\bibfnamefont {C.}~\bibnamefont {Ratti}},\ }\href {\doibase
  10.1103/PhysRevLett.111.202302} {\bibfield  {journal} {\bibinfo  {journal}
  {Phys. Rev. Lett.}\ }\textbf {\bibinfo {volume} {111}},\ \bibinfo {pages}
  {202302} (\bibinfo {year} {2013})},\ \Eprint {http://arxiv.org/abs/1305.6297}
  {arXiv:1305.6297 [hep-lat]} \BibitemShut {NoStop}%
\bibitem [{\citenamefont {Chatterjee}\ \emph {et~al.}(2017)\citenamefont
  {Chatterjee}, \citenamefont {Dash},\ and\ \citenamefont
  {Mohanty}}]{Chatterjee:2016cog}%
  \BibitemOpen
  \bibfield  {author} {\bibinfo {author} {\bibfnamefont {S.}~\bibnamefont
  {Chatterjee}}, \bibinfo {author} {\bibfnamefont {A.~K.}\ \bibnamefont
  {Dash}}, \ and\ \bibinfo {author} {\bibfnamefont {B.}~\bibnamefont
  {Mohanty}},\ }\href {\doibase 10.1088/1361-6471/aa8857} {\bibfield  {journal}
  {\bibinfo  {journal} {J. Phys.}\ }\textbf {\bibinfo {volume} {G44}},\
  \bibinfo {pages} {105106} (\bibinfo {year} {2017})},\ \Eprint
  {http://arxiv.org/abs/1608.00643} {arXiv:1608.00643 [nucl-th]} \BibitemShut
  {NoStop}%
\bibitem [{\citenamefont {Bellwied}(2016)}]{Bellwied:2016kpj}%
  \BibitemOpen
  \bibfield  {author} {\bibinfo {author} {\bibfnamefont {R.}~\bibnamefont
  {Bellwied}},\ }\bibfield  {booktitle} {\emph {\bibinfo {booktitle}
  {{Proceedings, 32th Winter Workshop on Nuclear Dynamics (WWND 2016):
  Guadeloupe, French West Indies, February 26-March 5, 2016}}},\ }\href
  {\doibase 10.1088/1742-6596/736/1/012018} {\bibfield  {journal} {\bibinfo
  {journal} {J. Phys. Conf. Ser.}\ }\textbf {\bibinfo {volume} {736}},\
  \bibinfo {pages} {012018} (\bibinfo {year} {2016})}\BibitemShut {NoStop}%
\bibitem [{\citenamefont {Alba}\ \emph {et~al.}(2018)\citenamefont {Alba},
  \citenamefont {Vovchenko}, \citenamefont {Gorenstein},\ and\ \citenamefont
  {Stoecker}}]{Alba:2016hwx}%
  \BibitemOpen
  \bibfield  {author} {\bibinfo {author} {\bibfnamefont {P.}~\bibnamefont
  {Alba}}, \bibinfo {author} {\bibfnamefont {V.}~\bibnamefont {Vovchenko}},
  \bibinfo {author} {\bibfnamefont {M.~I.}\ \bibnamefont {Gorenstein}}, \ and\
  \bibinfo {author} {\bibfnamefont {H.}~\bibnamefont {Stoecker}},\ }\href
  {\doibase 10.1016/j.nuclphysa.2018.03.007} {\bibfield  {journal} {\bibinfo
  {journal} {Nucl. Phys.}\ }\textbf {\bibinfo {volume} {A974}},\ \bibinfo
  {pages} {22} (\bibinfo {year} {2018})},\ \Eprint
  {http://arxiv.org/abs/1606.06542} {arXiv:1606.06542 [hep-ph]} \BibitemShut
  {NoStop}%
\bibitem [{\citenamefont {Vovchenko}\ \emph {et~al.}(2018)\citenamefont
  {Vovchenko}, \citenamefont {Gorenstein},\ and\ \citenamefont
  {Stoecker}}]{Vovchenko:2018fmh}%
  \BibitemOpen
  \bibfield  {author} {\bibinfo {author} {\bibfnamefont {V.}~\bibnamefont
  {Vovchenko}}, \bibinfo {author} {\bibfnamefont {M.~I.}\ \bibnamefont
  {Gorenstein}}, \ and\ \bibinfo {author} {\bibfnamefont {H.}~\bibnamefont
  {Stoecker}},\ }\href {\doibase 10.1103/PhysRevC.98.034906} {\bibfield
  {journal} {\bibinfo  {journal} {Phys. Rev.}\ }\textbf {\bibinfo {volume}
  {C98}},\ \bibinfo {pages} {034906} (\bibinfo {year} {2018})},\ \Eprint
  {http://arxiv.org/abs/1807.02079} {arXiv:1807.02079 [nucl-th]} \BibitemShut
  {NoStop}%
\bibitem [{\citenamefont {Dash}\ \emph
  {et~al.}(2018{\natexlab{a}})\citenamefont {Dash}, \citenamefont {Samanta},\
  and\ \citenamefont {Mohanty}}]{Dash:2018mep}%
  \BibitemOpen
  \bibfield  {author} {\bibinfo {author} {\bibfnamefont {A.}~\bibnamefont
  {Dash}}, \bibinfo {author} {\bibfnamefont {S.}~\bibnamefont {Samanta}}, \
  and\ \bibinfo {author} {\bibfnamefont {B.}~\bibnamefont {Mohanty}},\
  }\href@noop {} {\  (\bibinfo {year} {2018}{\natexlab{a}})},\ \Eprint
  {http://arxiv.org/abs/1806.02117} {arXiv:1806.02117 [hep-ph]} \BibitemShut
  {NoStop}%
\bibitem [{\citenamefont {Andronic}\ \emph {et~al.}(2018)\citenamefont
  {Andronic}, \citenamefont {Braun-Munzinger}, \citenamefont {Friman},
  \citenamefont {Lo}, \citenamefont {Redlich},\ and\ \citenamefont
  {Stachel}}]{Andronic:2018qqt}%
  \BibitemOpen
  \bibfield  {author} {\bibinfo {author} {\bibfnamefont {A.}~\bibnamefont
  {Andronic}}, \bibinfo {author} {\bibfnamefont {P.}~\bibnamefont
  {Braun-Munzinger}}, \bibinfo {author} {\bibfnamefont {B.}~\bibnamefont
  {Friman}}, \bibinfo {author} {\bibfnamefont {P.~M.}\ \bibnamefont {Lo}},
  \bibinfo {author} {\bibfnamefont {K.}~\bibnamefont {Redlich}}, \ and\
  \bibinfo {author} {\bibfnamefont {J.}~\bibnamefont {Stachel}},\ }\href@noop
  {} {\  (\bibinfo {year} {2018})},\ \Eprint {http://arxiv.org/abs/1808.03102}
  {arXiv:1808.03102 [hep-ph]} \BibitemShut {NoStop}%
\bibitem [{\citenamefont {Dash}\ \emph
  {et~al.}(2018{\natexlab{b}})\citenamefont {Dash}, \citenamefont {Samanta},\
  and\ \citenamefont {Mohanty}}]{Dash:2018can}%
  \BibitemOpen
  \bibfield  {author} {\bibinfo {author} {\bibfnamefont {A.}~\bibnamefont
  {Dash}}, \bibinfo {author} {\bibfnamefont {S.}~\bibnamefont {Samanta}}, \
  and\ \bibinfo {author} {\bibfnamefont {B.}~\bibnamefont {Mohanty}},\ }\href
  {\doibase 10.1103/PhysRevC.97.055208} {\bibfield  {journal} {\bibinfo
  {journal} {Phys. Rev.}\ }\textbf {\bibinfo {volume} {C97}},\ \bibinfo {pages}
  {055208} (\bibinfo {year} {2018}{\natexlab{b}})},\ \Eprint
  {http://arxiv.org/abs/1802.04998} {arXiv:1802.04998 [nucl-th]} \BibitemShut
  {NoStop}%
\bibitem [{\citenamefont {Wheaton}\ \emph {et~al.}(2009)\citenamefont
  {Wheaton}, \citenamefont {Cleymans},\ and\ \citenamefont
  {Hauer}}]{Wheaton:2004qb}%
  \BibitemOpen
  \bibfield  {author} {\bibinfo {author} {\bibfnamefont {S.}~\bibnamefont
  {Wheaton}}, \bibinfo {author} {\bibfnamefont {J.}~\bibnamefont {Cleymans}}, \
  and\ \bibinfo {author} {\bibfnamefont {M.}~\bibnamefont {Hauer}},\ }\href
  {\doibase 10.1016/j.cpc.2008.08.001} {\bibfield  {journal} {\bibinfo
  {journal} {Comput. Phys. Commun.}\ }\textbf {\bibinfo {volume} {180}},\
  \bibinfo {pages} {84} (\bibinfo {year} {2009})},\ \Eprint
  {http://arxiv.org/abs/hep-ph/0407174} {arXiv:hep-ph/0407174 [hep-ph]}
  \BibitemShut {NoStop}%
\bibitem [{Note1()}]{Note1}%
  \BibitemOpen
  \bibinfo {note} {B. Hippolyte and Y. Schutz,
  https://github.com/thermus-project/THERMUS}\BibitemShut {NoStop}%
\bibitem [{\citenamefont {Braun-Munzinger}\ \emph {et~al.}(2002)\citenamefont
  {Braun-Munzinger}, \citenamefont {Cleymans}, \citenamefont {Oeschler},\ and\
  \citenamefont {Redlich}}]{BraunMunzinger:2001as}%
  \BibitemOpen
  \bibfield  {author} {\bibinfo {author} {\bibfnamefont {P.}~\bibnamefont
  {Braun-Munzinger}}, \bibinfo {author} {\bibfnamefont {J.}~\bibnamefont
  {Cleymans}}, \bibinfo {author} {\bibfnamefont {H.}~\bibnamefont {Oeschler}},
  \ and\ \bibinfo {author} {\bibfnamefont {K.}~\bibnamefont {Redlich}},\ }\href
  {\doibase 10.1016/S0375-9474(01)01257-X} {\bibfield  {journal} {\bibinfo
  {journal} {Nucl. Phys.}\ }\textbf {\bibinfo {volume} {A697}},\ \bibinfo
  {pages} {902} (\bibinfo {year} {2002})},\ \Eprint
  {http://arxiv.org/abs/hep-ph/0106066} {arXiv:hep-ph/0106066 [hep-ph]}
  \BibitemShut {NoStop}%
\bibitem [{\citenamefont {Letessier}\ \emph {et~al.}(1995)\citenamefont
  {Letessier}, \citenamefont {Tounsi}, \citenamefont {Heinz}, \citenamefont
  {Sollfrank},\ and\ \citenamefont {Rafelski}}]{Letessier:1993hi}%
  \BibitemOpen
  \bibfield  {author} {\bibinfo {author} {\bibfnamefont {J.}~\bibnamefont
  {Letessier}}, \bibinfo {author} {\bibfnamefont {A.}~\bibnamefont {Tounsi}},
  \bibinfo {author} {\bibfnamefont {U.~W.}\ \bibnamefont {Heinz}}, \bibinfo
  {author} {\bibfnamefont {J.}~\bibnamefont {Sollfrank}}, \ and\ \bibinfo
  {author} {\bibfnamefont {J.}~\bibnamefont {Rafelski}},\ }\href {\doibase
  10.1103/PhysRevD.51.3408} {\bibfield  {journal} {\bibinfo  {journal} {Phys.
  Rev.}\ }\textbf {\bibinfo {volume} {D51}},\ \bibinfo {pages} {3408} (\bibinfo
  {year} {1995})},\ \Eprint {http://arxiv.org/abs/hep-ph/9212210}
  {arXiv:hep-ph/9212210 [hep-ph]} \BibitemShut {NoStop}%
\bibitem [{\citenamefont {Adam}\ \emph {et~al.}(2017)\citenamefont {Adam} \emph
  {et~al.}}]{ALICE:2017jyt}%
  \BibitemOpen
  \bibfield  {author} {\bibinfo {author} {\bibfnamefont {J.}~\bibnamefont
  {Adam}} \emph {et~al.} (\bibinfo {collaboration} {ALICE}),\ }\href {\doibase
  10.1038/nphys4111} {\bibfield  {journal} {\bibinfo  {journal} {Nature Phys.}\
  }\textbf {\bibinfo {volume} {13}},\ \bibinfo {pages} {535} (\bibinfo {year}
  {2017})},\ \Eprint {http://arxiv.org/abs/1606.07424} {arXiv:1606.07424
  [nucl-ex]} \BibitemShut {NoStop}%
\bibitem [{\citenamefont {Abelev}\ \emph
  {et~al.}(2014{\natexlab{a}})\citenamefont {Abelev} \emph
  {et~al.}}]{Abelev:2013haa}%
  \BibitemOpen
  \bibfield  {author} {\bibinfo {author} {\bibfnamefont {B.~B.}\ \bibnamefont
  {Abelev}} \emph {et~al.} (\bibinfo {collaboration} {ALICE}),\ }\href
  {\doibase 10.1016/j.physletb.2013.11.020} {\bibfield  {journal} {\bibinfo
  {journal} {Phys. Lett.}\ }\textbf {\bibinfo {volume} {B728}},\ \bibinfo
  {pages} {25} (\bibinfo {year} {2014}{\natexlab{a}})},\ \Eprint
  {http://arxiv.org/abs/1307.6796} {arXiv:1307.6796 [nucl-ex]} \BibitemShut
  {NoStop}%
\bibitem [{\citenamefont {Adam}\ \emph {et~al.}(2016)\citenamefont {Adam} \emph
  {et~al.}}]{Adam:2015vsf}%
  \BibitemOpen
  \bibfield  {author} {\bibinfo {author} {\bibfnamefont {J.}~\bibnamefont
  {Adam}} \emph {et~al.} (\bibinfo {collaboration} {ALICE}),\ }\href {\doibase
  10.1016/j.physletb.2016.05.027} {\bibfield  {journal} {\bibinfo  {journal}
  {Phys. Lett.}\ }\textbf {\bibinfo {volume} {B758}},\ \bibinfo {pages} {389}
  (\bibinfo {year} {2016})},\ \Eprint {http://arxiv.org/abs/1512.07227}
  {arXiv:1512.07227 [nucl-ex]} \BibitemShut {NoStop}%
\bibitem [{\citenamefont {Abelev}\ \emph
  {et~al.}(2013{\natexlab{a}})\citenamefont {Abelev} \emph
  {et~al.}}]{Abelev:2013vea}%
  \BibitemOpen
  \bibfield  {author} {\bibinfo {author} {\bibfnamefont {B.}~\bibnamefont
  {Abelev}} \emph {et~al.} (\bibinfo {collaboration} {ALICE}),\ }\href
  {\doibase 10.1103/PhysRevC.88.044910} {\bibfield  {journal} {\bibinfo
  {journal} {Phys. Rev.}\ }\textbf {\bibinfo {volume} {C88}},\ \bibinfo {pages}
  {044910} (\bibinfo {year} {2013}{\natexlab{a}})},\ \Eprint
  {http://arxiv.org/abs/1303.0737} {arXiv:1303.0737 [hep-ex]} \BibitemShut
  {NoStop}%
\bibitem [{\citenamefont {Abelev}\ \emph
  {et~al.}(2013{\natexlab{b}})\citenamefont {Abelev} \emph
  {et~al.}}]{Abelev:2013xaa}%
  \BibitemOpen
  \bibfield  {author} {\bibinfo {author} {\bibfnamefont {B.~B.}\ \bibnamefont
  {Abelev}} \emph {et~al.} (\bibinfo {collaboration} {ALICE}),\ }\href
  {\doibase 10.1103/PhysRevLett.111.222301} {\bibfield  {journal} {\bibinfo
  {journal} {Phys. Rev. Lett.}\ }\textbf {\bibinfo {volume} {111}},\ \bibinfo
  {pages} {222301} (\bibinfo {year} {2013}{\natexlab{b}})},\ \Eprint
  {http://arxiv.org/abs/1307.5530} {arXiv:1307.5530 [nucl-ex]} \BibitemShut
  {NoStop}%
\bibitem [{\citenamefont {Abelev}\ \emph
  {et~al.}(2014{\natexlab{b}})\citenamefont {Abelev} \emph
  {et~al.}}]{Abelev:2013zaa}%
  \BibitemOpen
  \bibfield  {author} {\bibinfo {author} {\bibfnamefont {B.~B.}\ \bibnamefont
  {Abelev}} \emph {et~al.} (\bibinfo {collaboration} {ALICE}),\ }\href
  {\doibase 10.1016/j.physletb.2014.05.052, 10.1016/j.physletb.2013.11.048}
  {\bibfield  {journal} {\bibinfo  {journal} {Phys. Lett.}\ }\textbf {\bibinfo
  {volume} {B728}},\ \bibinfo {pages} {216} (\bibinfo {year}
  {2014}{\natexlab{b}})},\ \bibinfo {note} {[Erratum: Phys.
  Lett.B734,409(2014)]},\ \Eprint {http://arxiv.org/abs/1307.5543}
  {arXiv:1307.5543 [nucl-ex]} \BibitemShut {NoStop}%
\bibitem [{\citenamefont {Abelev}\ \emph {et~al.}(2007)\citenamefont {Abelev}
  \emph {et~al.}}]{Abelev:2006cs}%
  \BibitemOpen
  \bibfield  {author} {\bibinfo {author} {\bibfnamefont {B.~I.}\ \bibnamefont
  {Abelev}} \emph {et~al.} (\bibinfo {collaboration} {STAR}),\ }\href {\doibase
  10.1103/PhysRevC.75.064901} {\bibfield  {journal} {\bibinfo  {journal} {Phys.
  Rev.}\ }\textbf {\bibinfo {volume} {C75}},\ \bibinfo {pages} {064901}
  (\bibinfo {year} {2007})},\ \Eprint {http://arxiv.org/abs/nucl-ex/0607033}
  {arXiv:nucl-ex/0607033} \BibitemShut {NoStop}%
\bibitem [{\citenamefont {Becattini}\ \emph {et~al.}(2011)\citenamefont
  {Becattini}, \citenamefont {Castorina}, \citenamefont {Milov},\ and\
  \citenamefont {Satz}}]{Becattini:2009ee}%
  \BibitemOpen
  \bibfield  {author} {\bibinfo {author} {\bibfnamefont {F.}~\bibnamefont
  {Becattini}}, \bibinfo {author} {\bibfnamefont {P.}~\bibnamefont
  {Castorina}}, \bibinfo {author} {\bibfnamefont {A.}~\bibnamefont {Milov}}, \
  and\ \bibinfo {author} {\bibfnamefont {H.}~\bibnamefont {Satz}},\ }\href
  {\doibase 10.1088/0954-3899/38/2/025002} {\bibfield  {journal} {\bibinfo
  {journal} {J. Phys.}\ }\textbf {\bibinfo {volume} {G38}},\ \bibinfo {pages}
  {025002} (\bibinfo {year} {2011})},\ \Eprint {http://arxiv.org/abs/0912.2855}
  {arXiv:0912.2855 [hep-ph]} \BibitemShut {NoStop}%
\bibitem [{\citenamefont {Becattini}\ \emph {et~al.}(2010)\citenamefont
  {Becattini}, \citenamefont {Castorina}, \citenamefont {Milov},\ and\
  \citenamefont {Satz}}]{Becattini:2010sk}%
  \BibitemOpen
  \bibfield  {author} {\bibinfo {author} {\bibfnamefont {F.}~\bibnamefont
  {Becattini}}, \bibinfo {author} {\bibfnamefont {P.}~\bibnamefont
  {Castorina}}, \bibinfo {author} {\bibfnamefont {A.}~\bibnamefont {Milov}}, \
  and\ \bibinfo {author} {\bibfnamefont {H.}~\bibnamefont {Satz}},\ }\href
  {\doibase 10.1140/epjc/s10052-010-1265-y} {\bibfield  {journal} {\bibinfo
  {journal} {Eur. Phys. J.}\ }\textbf {\bibinfo {volume} {C66}},\ \bibinfo
  {pages} {377} (\bibinfo {year} {2010})},\ \Eprint
  {http://arxiv.org/abs/0911.3026} {arXiv:0911.3026 [hep-ph]} \BibitemShut
  {NoStop}%
\bibitem [{\citenamefont {Vislavicius}\ and\ \citenamefont
  {Kalweit}(2016)}]{Vislavicius:2016rwi}%
  \BibitemOpen
  \bibfield  {author} {\bibinfo {author} {\bibfnamefont {V.}~\bibnamefont
  {Vislavicius}}\ and\ \bibinfo {author} {\bibfnamefont {A.}~\bibnamefont
  {Kalweit}},\ }\href@noop {} {\  (\bibinfo {year} {2016})},\ \Eprint
  {http://arxiv.org/abs/nucl-ex/1610.03001} {arXiv:nucl-ex/1610.03001
  [nucl-ex]} \BibitemShut {NoStop}%
\bibitem [{\citenamefont {Sharma}\ \emph {et~al.}(2018)\citenamefont {Sharma},
  \citenamefont {Cleymans},\ and\ \citenamefont {Kumar}}]{Sharma:2018owb}%
  \BibitemOpen
  \bibfield  {author} {\bibinfo {author} {\bibfnamefont {N.}~\bibnamefont
  {Sharma}}, \bibinfo {author} {\bibfnamefont {J.}~\bibnamefont {Cleymans}}, \
  and\ \bibinfo {author} {\bibfnamefont {L.}~\bibnamefont {Kumar}},\ }\href
  {\doibase 10.1140/epjc/s10052-018-5767-3} {\bibfield  {journal} {\bibinfo
  {journal} {Eur. Phys. J.}\ }\textbf {\bibinfo {volume} {C78}},\ \bibinfo
  {pages} {288} (\bibinfo {year} {2018})},\ \Eprint
  {http://arxiv.org/abs/1802.07972} {arXiv:1802.07972 [hep-ph]} \BibitemShut
  {NoStop}%
\end{thebibliography}%
\end{document}